\documentclass[twocolumn,showpacs,preprintnumbers,amsmath,amssymb,floatfix,aps]{revtex4}
\usepackage{amsmath}
\usepackage{graphicx}
\usepackage{dcolumn}
\usepackage{bm}

\usepackage{graphicx}

\begin{document}

\title{Vortex Fractionalization in a Josephson Ladder}

\date{\today}

\begin{abstract}

We show numerically that, in a Josephson ladder with periodic
boundary conditions and subject to a suitable transverse magnetic
field,
 a vortex excitation can spontaneously break up into
 two or more fractional excitations.  If the ladder
 has N plaquettes, and N is divisible by an integer
 q, then in an applied transverse field of 1/q flux
 quanta per plaquette the ground state is a regular
 pattern of one fluxon every q plaquettes.  When one
 additional fluxon is added to the ladder, it breaks
 up into q fractional fluxons, each carrying 1/q
 units  of vorticity.  The fractional fluxons are
basically walls between different domains of the ground state of the
underlying 1/q lattice.  The fractional fluxons are all
 depinned at the same applied current and move as a
unit.   For certain applied fields and ladder
 lengths, we show that there are isolated fractional
fluxons.  It is shown that the fractional fluxons would produce a
time-averaged voltage related in a characteristic way to the ac
voltage frequency.

\end{abstract}

\author{I. Tornes~\cite{email1}} \author{D. Stroud~\cite{email2}}

\affiliation{Department of Physics, The Ohio State University,Columbus, Ohio 43210}


\maketitle

\section{Introduction}

Ladder arrays of Josephson junctions have been extensively studied
both theoretically and experimentally (see, for example, Refs.
~\cite{binder,floria,kardar,binder1,flach,mazo,denniston,trias,granato,mazo1,caputo,ryu,
barahona,trees,yu}). They are of interest, in part, because they
show a rich variety of equilibrium and dynamical behavior.  In
addition, they are an important testing ground for concepts of phase
transitions and quantum behavior in lower dimensionality.  The
interest in such materials is even greater now, because advances in
microfabrication allow such ladders to be made almost to
specifications, in order to test theoretical predictions or possibly
to make new types of Josephson devices.

Recently Chandran and Kulkarni\cite{chandran} considered the
behavior of a hypothetical Josephson ladder array with alternating
$\pi$ and $0$ junctions.  A $\pi$ junction is one in which the
Josephson supercurrent $I = I_c\sin(\Delta\phi + \pi)$, where
$\Delta\phi$ is the phase difference across the junction and $I_c >
0$.  Such junctions can be formed in a variety of ways, e.g., by
controlling the energy distribution of the current-carrying states
in the normal metal within the junction\cite{baselmans}, by
preparing a junction between two cuprate superconductors in the
presence of bound states in the interface material\cite{lofwander},
or by connecting two superconductors across a ferromagnetic
layer\cite{ryazanov}. Thus, an array of alternating $0$ and $\pi$
junctions could, in principle, be made in the laboratory.   In Ref.\
\cite{chandran}, it was shown that when a $2\pi$ fluxon is
introduced into such a $0-\pi$ ladder, it will break up into two
separate $\pi$ fluxons, each of magnitude $\Phi_0/2$. This
fractionalized vortex was predicted to have unusual current-voltage
characteristics which could readily be detected experimentally.

In this paper, we show that similar fractionalized vortices are
produced in {\em conventional} ladder Josephson arrays made of $0$
junctions, in a transverse magnetic field.  Specifically, we
consider a ladder having $N$ square plaquettes, where $N$ is
divisible by an integer $q$, in a transverse magnetic field.  If
that field is of magnitude equal to one flux quantum per $q$
plaquettes, the ground state of the ladder is a periodic array of
$N/q$ equally spaced fluxons.  If now one fluxon is added to that
ladder, we find that the added fluxon breaks up into $q$ fractional
fluxons, each of magnitude $\Phi_0/q$.  If $q = 2$, the fluxon
pattern is similar to that found in Ref.\ \cite{chandran}, but we
find analogous patterns for all other values of $q$ which we have
tested. These extra fluxons have $IV$ characteristics and vorticity
patterns which should be readily detectable experimentally.
Furthermore, the fractionalized fluxons are typically "confined"
that is, they are typically depinned at the same current and move
at the same velocity. For an array in which $N$ is not divisible by
an integer $q$, we show numerically that it is possible to produce
{\em isolated} fractional fluxons, by applying a transverse magnetic
field of suitable magnitude.

The remainder of this paper is arranged as follows. In Section II,
we develop the equations of motion from a standard Lagrangian for
the Josephson ladder. In Section III, we give our numerical results,
which show that fractional vortices are generated in the ladder at
suitable applied transverse fields. Finally, in Section IV, we
briefly compare our results to other models which give rise to
fractional excitations, and give a concluding discussion.

\section{Formalism}

\begin{figure}
\begin{center}
\includegraphics[width=0.45\textwidth]{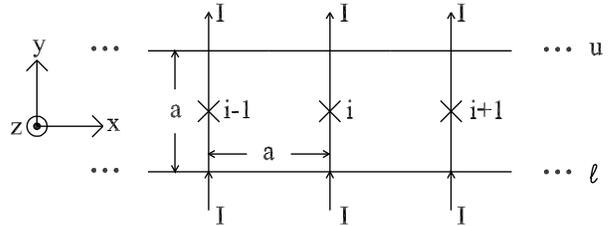} \caption
{\small Schematic of the ladder modeled in this paper.  We show the
$i-1$, $i$, and $i+1$ rungs.  $u$ and $\ell$ denote the upper and
lower edges of the ladder.  An external magnetic field is applied in
the ${\bf \hat{z}}$ direction; a dc driving current $I$ is applied to
the lower end of each rung in the $\hat{y}$ direction and extracted
from the upper end.  The ladder is assumed to have the topology of a
ring (periodic boundary conditions).  The full lines along the upper
and lower edges of the ladder (parallel to x) are treated simply as
inductors in our model.}\label{jos_ladder}
\end{center}
\end{figure}

We consider a Josephson ladder with periodic boundary conditions, as
shown schematically in Fig.\ \ref{jos_ladder}.  The ladder is
assumed to consist of a collection of small junctions, each of
critical current $I_c$, which are inductively coupled together. The
edges of the ladder are parallel to the $x$ axis, while the rungs
are in the $y$ direction. A dc current $I$ is injected into each
junction on one edge of the ladder and extracted from each junction
on the other edge. A magnetic field ${\bf B} = B{\bf \hat{z}}$ is
applied perpendicular to the ladder. The geometry is readily
achievable experimentally, in the form of a coplanar ring
perpendicular to ${\bf B}$.

In the absence of dissipation, it is convenient to describe the
ladder by the following Lagrangian:
\begin{equation}
{\cal L} = K - V
\end{equation}
where
\begin{equation}
K = \sum_{i = 1}^N\frac{\hbar^2}{2U}(\dot{\theta}_{i,u} -
\dot{\theta}_{i,\ell})^2
\end{equation}
and
\begin{eqnarray}
V &=& \sum_{i=1}^N[-J_1\cos(\theta_{i,\ell}- \theta_{i,u}- A_i)+
J_2(\theta_{i,\ell}- \theta_{i+1,\ell})^2\nonumber \\
&+&J_2(\theta_{i,u}-\theta_{i+1,u})^2  -J_3(\theta_{i,u}
-\theta_{i,\ell} - A_i)].
\end{eqnarray}
Here $K$ represents the charging energy of the small junctions, $U =
4e^2/C$, and $C$ is the capacitance of the small junction. $J_1 =
\hbar I_c/2e$ is the Josephson coupling energy of the small
junction, $J_3 = \hbar I/(2e)$, and $J_2 = \hbar^2/(8e^2L)$, where
$L$ is the inductance of the wire segments coupling points
($i,\ell$) to ($i+1,\ell$) and ($i,u$) to ($i+1,u$). Finally, if we
use the gauge ${\bf A} = Bx{\bf \hat{y}}$, where ${\bf A}$ is the
vector potential, then $A_i = (2\pi/\Phi_0)\int_{i,\ell}^{i,u} {\bf
A}\cdot{\bf dl} = (2\pi/\Phi_0)a^2iB$, where $\Phi_0 = hc/(2e)$ is
the flux quantum, $a$ is the lattice constant (see Fig.\ 1), and the
subscripts $\ell$ and $u$ indicate the lower and upper ends of the
junction, as shown in the Figure. The dots indicate time
derivatives.  Note that we neglect the self-induced magnetic field,
by taking $B$ equal to the externally applied field.

If we now make the substitutions $\phi_i = \theta_{i,u} -
\theta_{i,\ell}$, $\chi_i =
(\theta_{i,\ell}+\theta_{i,u})$\cite{kardar,ryu}, we may reexpress
the Lagrangian as
\begin{eqnarray}
{\cal L} = \sum_{i=1}^N\bigg[\frac{E_C}{2}\dot{\phi}_i^2
+ J_1\cos(\phi_i-A_i) +J_3\phi_i + \nonumber \\
-\frac{J_2}{2}[(\chi_i-\chi_{i+1})^2 +(\phi_i- \phi_{i+1})^2]\bigg]
\end{eqnarray}
The Lagrange equations of motion are then
$(d/dt(\partial{\cal L}/\partial\dot{\phi}_i)) = \partial{\cal
L}/\partial \phi$, with an analogous equation for the variables
$\chi_i$.  Since ${\cal L}$ is independent of $\dot{\chi}_i$, the
equation of motion for $\chi_i$ leads to the condition that $\chi_i-
\chi_{i+1}$ is independent of $i$.  After some algebra, it is
readily shown that the equation of motion for $\phi_i$ is
\begin{eqnarray}
\ddot{\phi}_i = &-& \sin(\phi_i + 2\pi i f) +
\lambda_J^2(\phi_{i-1}-2\phi_i+\phi_{i+1}) \nonumber \\
&+& (I/I_c)\phi_i -\dot{\phi}_i/Q_J.
\end{eqnarray}
Here, we have introduced the frustration $f = Ba^2/\Phi_0$, and a
dimensionless time $\tau = \omega_pt$, where $\omega_p =
[2eI_c/(\hbar C)]^{1/2}$ is the junction plasma frequency, and
$\lambda_J = (J_2/J_1)^{1/2}$ is a dimensionless Josephson
penetration depth. The dot now represents a derivative with respect
to $\tau$.  Finally, we have added an additional dissipative term
$-\dot{\phi}_i/Q_J$ by hand to the equation of motion, where $Q_J$
is a dimensionless junction quality factor, assumed to be the same
for all junctions.  In the resistively and capacitively shunted
junction model (RCSJ)\cite{rcsj}, $Q_J = (2eR^2I_cC/\hbar)^{1/2}$,
where $R$ is the junction shunt resistance.

\section{Numerical Results}

\subsection{Numerical Method}

In most of our calculations, we have studied a ladder with $N = 120$
rungs and periodic boundary conditions.  This number of rungs was
chosen because it is the smallest integer divisible by $2$, $3$,
$4$, $5$ and $6$, and is sufficiently large to allow the
fractionalization to be apparent in the simulations.  We solved our
set of $240$ first order, non-linear, differential equations using a
constant-time-step, fourth-order Runge-Kutta method~\cite{num_rec}.
We have generally considered underdamped junctions, with $Q_J=10$,
and values of $\lambda_J^2$ between $0.01$ and $10$.  We initialized
the system in a random phase configuration at $\tau = 0$ and $I/I_c
= 0$.  The equations are integrated forward in time up to $\tau_{eq}
= 3000$, by which time we assume the system has reached a steady
state.  We then calculate the time-averaged voltage drop across a
rung of the ladder (also averaged over the rungs) by averaging over
an additional $\tau_{max} - \tau_{eq}$ dimensionless time units, i.
e.,
\begin{equation}
\frac{<V>}{RI_c} = \frac{1}{NQ_J(\tau_{max}-\tau_{eq})} \sum_{i =
1}^N \int_{\tau_{eq}}^{\tau_{max}} \frac{d\,\phi_i}{d\,\tau}
d\,\tau,
\end{equation}
where $\tau_{max} = 5000$.
The dc driving current is then increased by $0.01 I_c$ and the
calculation is repeated.  We continue ramping up the current by
steps of $0.01$ until $I/I_c = 1.2$, then decrease $I/I_c$ in steps
of $0.01$ back to zero. These calculations give the IV curves shown
below in Figs.\ ~\ref{f_0p5}(a), \ref{f_61over120}, \ref{f_1over3}
 and \ref{f_41over120}.

\begin{figure}
\begin{center}
\includegraphics[width=0.45\textwidth]{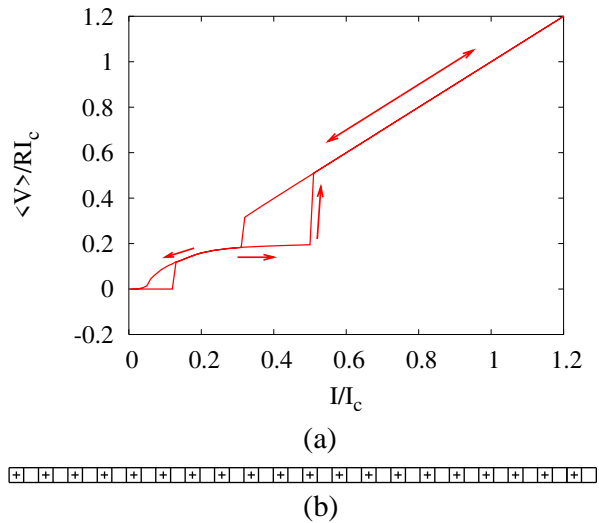}
\caption { \small (a) IV curves for $f = 1/2$, $Q_J = 10.0$, and
$\lambda_J^2 = 1.0$ for increasing and decreasing current (the
directions are indicated by arrows).  There are two voltage jumps in
the IV curve on the increasing branch: at the first, the fluxon
lattice is depinned; at the second, the junctions switch to the
voltage state. (b) Schematic of ground state vortex number
configuration for the $f = 1/2$ ladder at $I/I_c = 0$; the $+$'s and
blank squares represent vortex numbers $n = 1$ and $n = 0$.  Only $40$
plaquettes are shown.}\label{f_0p5}
\end{center}
\end{figure}

\subsection{f = 1/2}

In Fig.\ ~\ref{f_0p5}(a), we show the full IV curve for $f = 1/2$,
$Q_J = 10$, and $\lambda_J^2 = 1$.  The directions of the arrows
indicate whether the current is increasing or decreasing.  The IV
curve is clearly hysteretic for this value of $Q_J$, and has two
discontinuous jumps on the increasing current branch.  At the lower
jump, near $I/I_c = 0.13$, the system jumps into a state where the
fluxon lattice is depinned and starts to move through the ladder as a
unit, giving rise to a finite voltage across the ladder. At the upper
jump, near $I/I_c = 0.5$, all the junctions switch to a finite voltage
state and the fluxon excitations are expelled from the ladder.

To represent the fluxon lattice pictorially, we use the concept of a
{\em vortex number}\cite{number}.  The vortex number of the
$\alpha^{th}$ plaquette is defined as
\begin{equation}
n_\alpha = f + \frac{1}{2\pi}\sum_{\mathrm{plaquette}}(\theta_i -
\theta_j - A_{ij}). \label{eq:vnum}
\end{equation}
Here the gauge-invariant phase difference for each leg of the
plaquette is written $(\theta_i-\theta_j-A_{ij})$ and is defined to
lie in the range $[-\pi, \pi]$; the summation is taken around the
plaquette in the counterclockwise direction (viewed from the
positive $z$ axis).  Thus $n_\alpha$ is $0$ or a positive or
negative integer for each plaquette. In the lowest energy state, we
expect that $\sum_\alpha n_\alpha = Nf$, where $N$ is the number of
plaquettes, and also that all the individual $n_\alpha = 0$ or $1$.

The two values of $n_\alpha$ can generally be distinguished by the
current pattern around the plaquette: $n_\alpha = 1$ and $0$
correspond to a clockwise and counterclockwise current flow,
respectively.  The expected ground state for $f = 1/2$ at $I/I_c =
0$ is shown pictorially in Fig.\ ~\ref{f_0p5}(b). The $+$'s and
empty squares represent plaquettes of $n_\alpha = 1$ and $0$,
respectively.

In practice, we have not found a way to calculate $n_\alpha$ directly,
because we do not compute the variables $\chi_i$ mentioned above.
Therefore, in order to confirm the vortex pattern at various fields
and currents, we calculate the "normalized flux," defined by the
relation $2\pi\Phi_i/\Phi_0 = \phi_i - \phi_{i+1} + 2\pi\,f$, in each
plaquette.  We show a number of examples of this normalized flux
below.

\begin{figure}
\begin{center}
\includegraphics[width=0.45\textwidth]{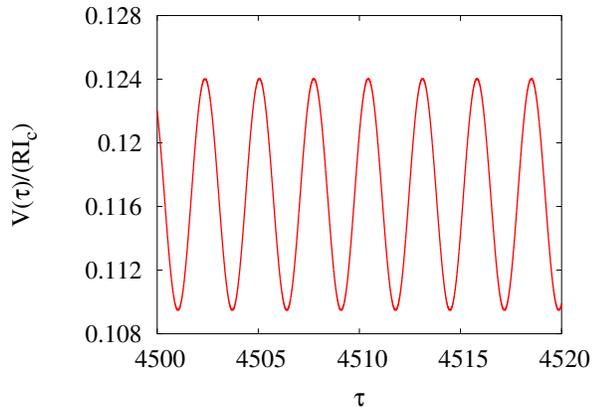}
\caption { \small Time-dependent voltage $V(\tau)/(RI_c)$ across one
rung of the ladder, averaged over the rungs, and plotted for
dimensionless time $\tau$ between $4500$ and $4520$, for $f = 1/2$
and the parameters of Fig.\ ~\ref{f_0p5}(a). The dimensionless
period of voltage oscillation is $\sim 2.7$ at $I/I_c =
0.13$.}\label{V_0p5}
\end{center}
\end{figure}

When the vortex lattice is depinned near $I/I_c = 0.13$, the entire
lattice of $+$'s moves as a unit through the ladder, generating a
voltage. In Fig.\ ~\ref{V_0p5}, we show the space-averaged voltage
\begin{equation}
\frac{V(\tau)}{RI_c} = \frac{1}{NQ_J}\sum_i^N \dot{\phi}_i(\tau).
\end{equation}
across the ladder for  $4500 < \tau <4520$, for the parameters of
Fig.\ ~\ref{f_0p5}(a) at $I/I_c = 0.13$, just above the lower jump
in the IV characteristics. We interpret the
periodic voltage oscillations as arising from the motion of the
fluxon lattice through the bumpy ``egg-carton'' potential provided
by the plaquettes of the Josephson ladder: the velocity of the
fluxon lattice varies periodically in time because the fluxon
lattice moves alternately faster and slower as it moves through the
steeper and less steep part of the egg carton\cite{lobb}.

To confirm this picture, we note from Fig.\ \ref{V_0p5} that the
dimensionless period of the voltage oscillation, which we denote
$\Delta \tau$, is $\sim 2.7$.   We expect that this should be the
dimensionless time required for the fluxon lattice to advance one
lattice constant $a$.  Now if the fluxon lattice moves a distance
$a$ during this period, then the average speed of the fluxon lattice
is $v = (Na)/T$, where $T$ is the time required for the lattice to
complete one circuit around the ladder. During this time, each
junction should experience a phase change of $2\pi\,N\,f$, since
$Nf$ fluxons cross each junction during one circuit. Thus, the
time-averaged voltage $\langle V \rangle$ across each junction
should be $[\hbar/(2e)]2\pi Nf/T$. Introducing a dimensionless
voltage $\langle V\rangle/(RI_c)$, and noting that the time required
to move one lattice constant $a$ is $\omega_p T/N =\Delta \tau$, we
obtain
\begin{equation}
\label{eq:voltage} \frac{\langle V\rangle}{RI_c} = \frac{2\pi
f}{Q_J\Delta \tau}.
\end{equation}
Using the parameters of Fig.\ ~\ref{f_0p5}(a), together with $I/I_c =
0.13$ and $\Delta\tau = 2.7$, eq.\ (\ref{eq:voltage}) gives $\langle
V\rangle/(RI_c) = 2\pi(1/2)/(10 \cdot 2.7) = 0.116$. The actual
time-averaged voltage read off from Fig.\ ~\ref{f_0p5}(a) at $I/I_c =
0.13$ is $0.117$, in excellent agreement with this calculated
value. Thus, the time-dependent voltage shown in Fig.\ \ref{V_0p5} is
consistent with the picture of a fluxon lattice moving as a unit
through the washboard potential provided by the ladder.  For currents
slightly higher than $I/I_c = 0.13$, this picture of a moving vortex
lattice continues to hold.  As the current is increased still further,
for $Q_J = 10$, the voltage is no longer perfectly periodic in time
(not shown in the Figure).  For $I/I_c$ greater than about $0.51$, the
individual junctions switch to a voltage state and the moving fluxon lattice
disappears.

\begin{figure}
\begin{center}
\includegraphics[width=0.45\textwidth]{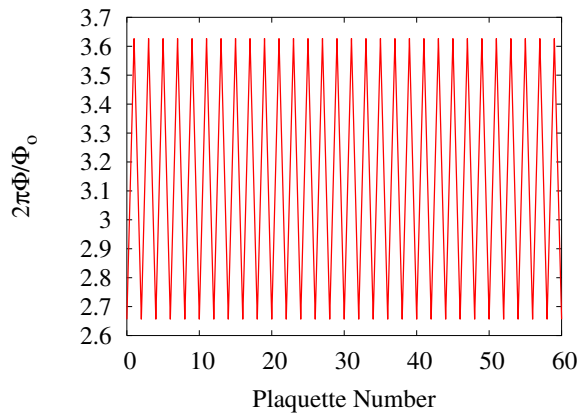}
\caption { \small Normalized "flux" $2\pi\Phi_i/\Phi_0$ (as defined
in the text) vs.\ plaquette number $i$, for the parameters of Fig.\
~\ref{f_0p5}(a) and $I/I_c = 0$.  For clarity, only $60$ of the
$120$ plaquettes are shown.}\label{flux_0p5}
\end{center}
\end{figure}

In Fig.\ ~\ref{flux_0p5}, we plot the calculated "normalized flux"
$\Phi_i/\Phi_0$ for $f = 1/2$ and $I/I_c = 0$.  Clearly,
$\Phi_i/\Phi_0$ alternates between two values, indicating that the
flux lattice is periodic with a period of two unit cells, as suggested
by Fig.\ 2(b). The two ground state fluxes are approximately $2\pi f
\pm f$ with $f = 1/2$.  A similar pattern has been seen previously in
Ref.\ ~\cite{chandran} for a ladder with alternating $0$ and $\pi$
junctions, and zero applied magnetic field.

\subsection{f = 1/2 +1/N}

\begin{figure}
\begin{center}
\includegraphics[width=0.45\textwidth]{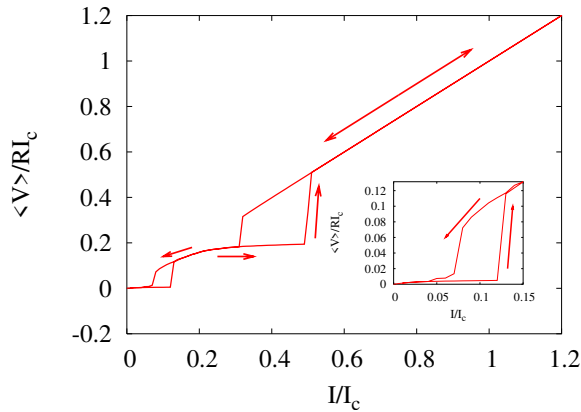}
\caption { \small IV curve for $f=61/120$, $Q_J = 10$ and $\lambda_J^2
  = 1$ for increasing and decreasing current (as indicated by arrows).  The inset shows the
  IV curve for
  $0<I/I_c<0.15$.  A finite voltage in this regime is produced by the two moving
  1/2-fluxons.}\label{f_61over120}
\end{center}
\end{figure}

\begin{figure}
\begin{center}
\includegraphics[width = 0.5\textwidth]{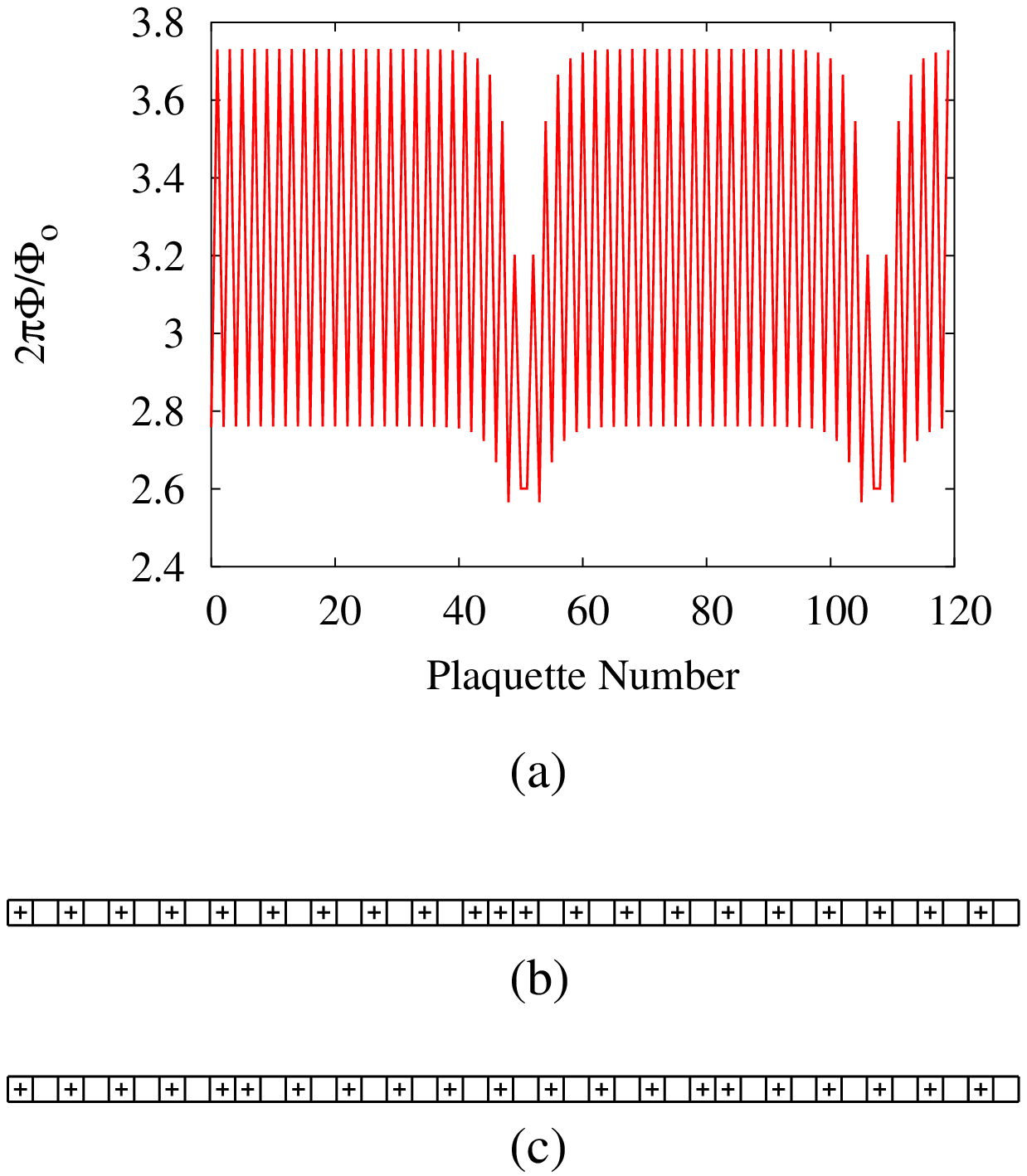}
\caption { \small (a) Normalized flux vs. plaquette number for the
  parameters of Fig.\ ~\ref{f_0p5}, except that $f=61/120$.
  (b) Schematic of an un-fractionalized, high-energy vortex
  charge configuration for $f = 61/120$.  (c) Schematic of the actual ground state vortex charge configuration at
  $f = 61/120$ and $I/I_c = 0$.  In both (b) and (c), we show only 40 of the 120 plaquettes.
  (In the actual ladder, the two fractional vortices are separated by 60 plaquettes.)}
  \label{flux_61over120}
\end{center}
\end{figure}

Next, we consider the case $f = 1/2 + 1/120$, equivalent to adding one
fluxon to the $f = 1/2$ ladder. The calculated results are shown in
Figs.\ ~\ref{f_61over120} and~\ref{flux_61over120}. Fig.\
~\ref{f_61over120} shows the IV characteristic as computed on
increasing and decreasing the current in steps of $0.01 I_c$. On the
increasing branch, there are now three steps in the IV curve.  Two of
the steps are easily seen from the Figure, but the lowest jump, which
occurs near $I/I_c = 0.01$, is less visible and shown in the inset for
$0 < I/I_c < 0.15$. The IV curve is hysteretic in two regions of
current, but non-hysteretic at the lowest non-zero voltages.  On the
decreasing branch there are still three distinct regimes of non-zero
voltage. We interpret these as follows: (i) at the lowest currents, a
single fluxon is depinned and moves through the fluxon lattice formed
by the remaining 60 fluxons; (ii) at higher currents, the fluxon
lattice is depinned, as at $f = 1/2$, generating a larger voltage; and
finally (iii) at still higher voltage, all the junctions switches to a
voltage state and the ladder jumps up to the resistive branch.

Fig.\ ~\ref{flux_61over120}(a) shows the flux pattern
$2\pi\Phi_i/\Phi_0$ for $f=61/120$ and $I/I_c = 0$, keeping the rest
of the parameters the same as in Fig.\ ~\ref{f_0p5}(a).  The $f = 1/2$
pattern is now distorted in {\em two} regimes, near $i = 50$ and $i =
110$, while in the rest of the ladder the pattern is similar to that
of $f = 1/2$.  To compare the two patterns more quantitatively, we
note that the sums $\sum_i\Phi_i/\Phi_0$ for Figs.\ ~\ref{flux_0p5}
and~\ref{flux_61over120}(a) are $60(2\pi)$ and $61(2\pi)$
respectively.  This result shows that the effect of the additional
magnetic field is to add one additional fluxon to the ladder.
However, this fluxon does not enter the ladder {\em as a unit}.
Instead, it is fractionalized into {\em two} 1/2-fluxons, each of
which carries 1/2 a unit of flux.  These two 1/2-fluxons obviously
repel one another, since they prefer to be situated as far away from
one another as possible, i. e. about 60 plaquettes apart in the
120-plaquette ladder.

To suggest intuitively why the extra fluxon fractionalizes in the
ladder, we show in Figs.\ \ref{flux_61over120} (b) and (c) two
possible scenarios for how the extra fluxon enters the ladder.  In
(b), the fluxon enters as a unit, without distorting the underlying
f = 1/2 fluxon lattice.  But this scenario leads to  $n = 1$
vortices in three consecutive plaquettes, as indicated by the three
adjacent plus signs. This is obviously a high energy state, since
vortices of like signs repel one another.   A more plausible state
is shown in Fig.\ \ref{flux_61over120} (c), where the extra fluxon
has fractionalized into two 1/2-fluxons. In this case, there are
only two points in the ladder where there are two adjacent $+$
signs, and none with three.   Our numerical simulations indicate
that this is indeed a lower energy state than the un-fractionalized
fluxon shown in Fig.\ \ref{flux_61over120} (b).  We have also found
that both halves of the fractionalized fluxon are depinned at the
same $I/I_c$, and move together.

Finally, we consider the expected time-dependent voltage for the two
moving 1/2-fluxons in the 61/120 ladder. As explained above, each
1/2 fluxon is really a kind of domain wall, consisting of two
adjacent $+$ vortex charges. The expected mechanism by which a
1/2-fluxon moves is that one of the two adjacent $+$ charges moves
to the nearest empty plaquette.  This jump causes the domain wall to
move by {\em two} plaquettes.  Thus, if the ladder has length $N$
plaquettes, a 1/2-fluxon will circulate around the ladder {\em
twice} in N moves. This causes a phase slip of $2\cdot\pi$, since
there is a $\pi$ phase slip each time the 1/2-fluxon circulates once
around the ladder.  Thus, if $T$ is the time required for the
1/2-fluxon to circulate once around the ladder, the time-averaged
voltage is $\langle V \rangle = (\hbar/2e)\pi/T$.  On the other
hand,  $V(t)$ for a single 1/2-fluxon should have an ac component
with a period of $1/\nu = 2T/N$, where $\nu$ is the ac frequency,
since $2T/N$ is the time required for a 1/2-fluxon to make a single
move of two plaquettes.

Combining these two relations, we get $\langle V \rangle =
[h/(2eN)]\nu$ for a single 1/2-fluxon.  This is precisely the
relation between $\langle V \rangle$ and $\nu$ which we find for a
ladder containing a single 1/2-fluxon, as is discussed later in this
paper.

For the present case of a $61/120$ ladder, there are, as already
mentioned, {\em two} 1/2-fluxons, as shown in Fig.\ 6.  Thus, at a
current $I$ in the regime where the {\em two} 1/2-fluxons are
moving, we expect that $\langle V \rangle$ will equal twice this
value, or
\begin{equation}
\langle V \rangle = 2[h/(2eN)]\nu,
\end{equation}
where $\nu$ represents the frequency of the ac voltage generated by
a the motion of a single 1/2-fluxon at the same current.  We can
rewrite this equation in terms of $Q_J$ and $\Delta \tau \equiv
\omega_p/\nu$ with the result
\begin{equation}
\frac{\langle V\rangle}{RI_c} = \frac{2\cdot(2\pi)}{NQ_J\Delta\tau}.
\label{eq:voltage2}
\end{equation}
In the present case $N = 120$ so the right-hand side equals
$2\pi/(60Q_J\Delta\tau)$.  The calculated $\langle V \rangle$, in
the regime where the two 1/2-fluxons are moving, agrees very well
with the prediction of eq.\ (\ref{eq:voltage2}).  Thus, the IV
characteristics in this regime are produced by two 1/2-fluxons
moving with the same average velocity.

\subsection{f = 1/3}

\begin{figure}
\begin{center}
\includegraphics[width=0.45\textwidth]{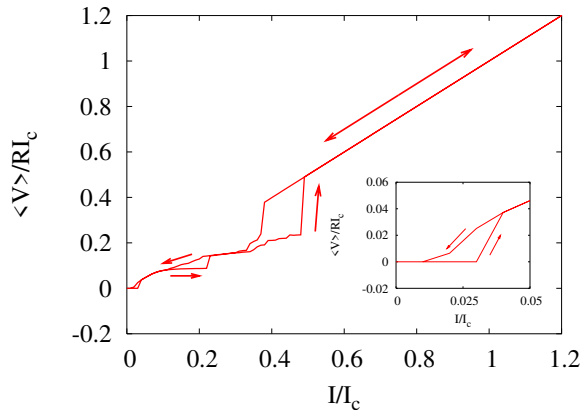}
\caption { \small IV curve for $f = 1/3$, $Q_J =
10.0$, $\lambda_J^2 =
  1.0$ and  $N=120$.  The inset shows the IV curve for $0<I/I_c<0.05$.}\label{f_1over3}
\end{center}
\end{figure}

\begin{figure}
\begin{center}
\includegraphics[width=0.45\textwidth]{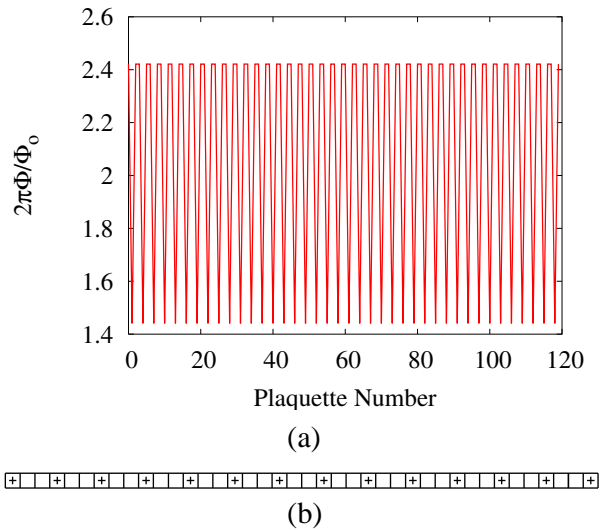}
\caption { \small (a) Normalized flux vs.\ plaquette number for the
parameters
  of Fig.\ ~\ref{f_1over3}. (b) Ground state
  vortex charge configuration for the same parameters as (a),  Only $40$
  of the $120$ plaquettes are shown.}\label{flux_1over3}
\end{center}
\end{figure}

If an added fluxon fractionalizes when added to a ladder at $f=1/2$,
what happens at other values of $f$?  To answer this question, we have
carried out similar calculations for $1/3,1/4,1/5$ and $1/6$ in an $N
= 120$ ladder, as well as for each of these values of $f$ with one
fluxon added or subtracted. The calculated IV characteristics and
normalized flux for $f=1/3$ are shown in Figs.\ ~\ref{f_1over3}
and~\ref{flux_1over3}(a).  The behavior of the IV curves shown in
Fig.\ ~\ref{f_1over3} is most easily understood on the decreasing
current branch.  For the highest currents, the junctions are all in
the voltage state, and the fluxon lattice has disappeared. At
intermediate currents, the $f = 1/3$ fluxon lattice is moving through
the washboard potential provided by the ladder.  At the lowest
currents, the fluxon lattice is pinned and the time-averaged voltage
vanishes. On the increasing current branch, the fluxon lattice appears
to depin in two stages, corresponding to the two voltage jumps visible
in the Figure.

Fig.\ ~\ref{flux_1over3}(a) shows the normalized flux
$\Phi_i/\Phi_0$ versus plaquette number $i$ at $I/I_c =0$ and
figure~\ref{flux_1over3}(b) shows the vortex charge pattern
configuration in the equilibrium state.  The flux pattern is again
periodic, as at $f = 1/2$, but now with a period of three
plaquettes, with fluxes $2\pi f + f$, $2\pi f + f$, and $2\pi f -
2f$.

 As a further check that the fluxon lattice really moves as a
unit, we have used eq.\ (\ref{eq:voltage}) to compare the
time-averaged voltage $\langle V\rangle$ at $I/I_c = 0.04$, on the
upward current sweep just above the first voltage jump, to the
dimensionless period $\Delta \tau$ of the a.\ c. component. The
prediction for $\langle V\rangle/(RI_c)$ from eq.\
(\ref{eq:voltage}) (namely $\langle V \rangle/(RI_c) = 0.0381$) is
in excellent agreement with the value calculated by a direct
time-average of $\langle V\rangle/(RI_c)$ ($0.0372$).

\begin{figure}
\begin{center}
\includegraphics[width=0.45\textwidth]{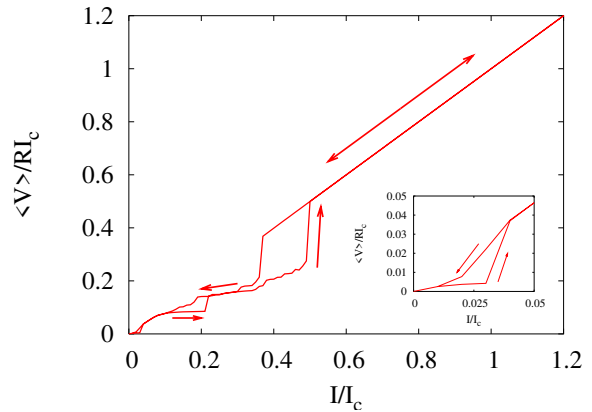}
\caption
{ \small IV curve for $f=41/120$, $Q_J = 10.0$ and
  $\lambda_J^2 = 1.0$, and $N = 120$.  There are three jumps on the increasing-current branch of
  the IV characteristics; the first occurs
  near $I/I_c = 0.01$, and the others can be easily seen in the
  Figure.  The inset shows the IV curve for
  $0<I/I_c<0.05$.  The interpretation of these regimes is given in the text.}\label{f_41over120}
\end{center}
\end{figure}

\begin{figure}
\begin{center}
\includegraphics[width=0.45\textwidth]{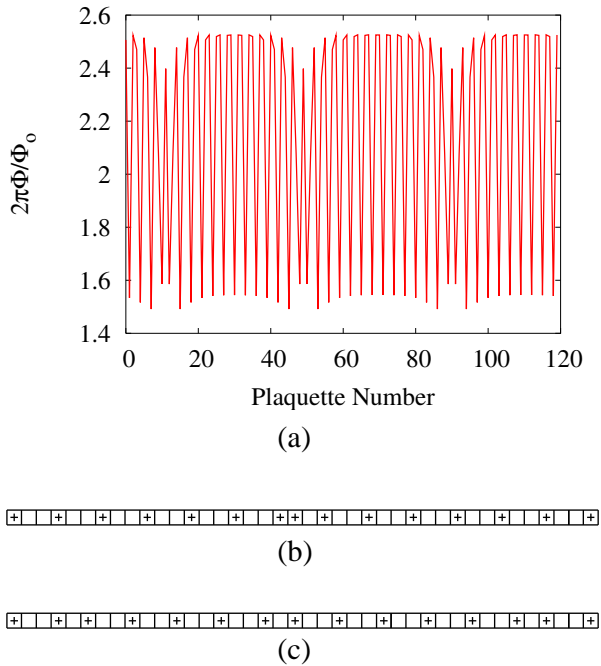}
\caption
{ \small (a) Normalized flux vs. plaquette number for the same
  parameters as in Fig.\ ~\ref{f_1over3}, except that
  $f=41/120$. (b) A possible high-energy vortex charge configuration
  for this field.  (c) Schematic of the actual ground-state vortex
  charge configuration, as obtained after iterating the equations of motion up to a time
  $\tau_{max} = 5000$  and $I/I_c = 0$.
  In both (b) and (c), we show only 40 of the 120 plaquettes.
  In the actual ladder at $f = 41/120$, the three 1/3-fluxons
  are separated by 40 plaquettes, as suggested in Fig.\ 10(a).}\label{flux_41over120}
\end{center}
\end{figure}

\subsection{f = 1/3 +1/N}

Next, we consider the behavior of the same system with one
additional fluxon.  Fig.\ ~\ref{f_41over120} shows the IV curve, and
Fig.\ ~\ref{flux_41over120}(a) displays the normalized flux vs.\
plaquette number at $I/I_c = 0$, for $f=1/3 +
1/120$.   In this case, the additional fluxon fractionalizes
into {\em three} excitations.  These are visible in Fig.\
~\ref{flux_41over120}(a) as regions which are distorted, in
comparison to the flux plot at $f = 1/3$.    As in Fig.\
~\ref{flux_61over120}(a), these regions show how the $f = 1/3$ flux
pattern changes to accommodate the additional flux quantum.  To
understand why this extra fluxon fractionalizes into three parts, we
show in Figs.\ \ref{flux_41over120}(b) and (c) two possible patterns
of vortex number when the additional fluxon is added.  The $f = 1/3$
ground state consists of a $+1$ vortex every three plaquettes.   If
the additional fluxon is un-fractionalized, we expect that it will
enter the ladder as in Fig.\ \ref{flux_41over120} (b).  This pattern
must have two neighboring plaquettes with a $+1$ vortex number, and
hence is a relatively high-energy state.  It is much more plausible
that the  lowest-energy pattern would never have $+1$ vortices in
adjacent plaquettes. Such a pattern is shown in Fig.\
\ref{flux_41over120} (c).  The $1/3$ fluxon shows up as a region
where two $+1$ vortices are separated by only one empty plaquette,
rather than two as in the unperturbed $f = 1/3$ pattern. To conserve
the number of $+1$ plaquettes, there must be three such regimes;
hence, the added fluxon is expected to fractionalize into three
1/3-fluxons, as we find numerically. These 1/3-fluxons are
equidistant in order to minimize the repulsive energy between these
added excitations.

When the extra fractionalized fluxon moves through the lattice, it
generates voltage.  Once again, we can use the analog of eq.\
(\ref{eq:voltage2}) to connect the time-averaged voltage to the
oscillation period of $V(\tau)$.  This analog is
\begin{equation}
\frac{\langle V \rangle}{RI_c} = \frac{3\cdot
(2\pi)}{NQ_J\Delta\tau}. \label{eq:voltage3}
\end{equation}
Here $\Delta\tau$ is now the period of the a.\ c.\ voltage in the
regime where three 1/3-fluxons are moving, measured in units of
$1/\omega_p$. When we use this relation to calculate  $\langle
V\rangle$ from our calculated $\Delta\tau$, at $I/I_c = 0.02$ (in
the regime of three moving 1/3-fluxons), we find that the value of
$\langle V \rangle$ obtained from eq.\ (\ref{eq:voltage3}), using
the calculated value of $\Delta\tau$ is in excellent agreement with
the value of $\langle V \rangle$ calculated directly.

\subsection{f = 1/q and f = 1/q + 1/N, with $q > 3$}

We have also carried out similar calculations for $1/4$, $1/5$ and
$1/6$.  The ground state (at $I/I_c = 0$) consists, as expected, of
a periodic array of $+1$ vortices separated by $q - 1$ zero vortices
(q = 4, 5 or 6).  When we carry out the corresponding calculations
for $f= 1/4 + 1/N$, $f = 1/5 + 1/N$ and $1/6 + 1/N$, we again find
that the added vortex fractionalizes into $q$ equally spaced pieces,
as suggested by our results for $q = 2$, $3$.

\subsection{Effects of Varying Ladder Length, Inductive Coupling,
and Damping Coefficient.  Effects of Removing One Fluxon.}

We have also studied $f = 1/q + 1/N$  ($q > 1$) for a ladder of
double the length, i. e., $240$ plaquettes. The flux patterns are
very similar to those of the shorter ladders, except that, of
course, the fractionalized fluxons are twice as far apart as for $N
= 120$; the IV curves are also qualitatively similar.

As another important test case, we have computed both the ground
state configurations and $IV$ characteristics for $f \rightarrow
f-1/N$, in order to compare the effects of {\em removing} rather
than adding one fluxon. (We do not display these results
graphically.) In general, the results are similar to those found for
adding one fluxon.  The results for $f = 1/2-1/N$ are nearly
identical to those for $f = 1/2+1/N$: the missing fluxon is
fractionalized into two half-fluxons. For $f = 1/q -1/N$, with $q>1$
an integer, the $IV$ and flux characteristics are similar to those
for $f = 1/q + 1/N$; however, the flux curves are not identical,
except for $q = 2$.  It is not surprising that one obtains different
results for adding and subtracting a fluxon for $q > 2$: only at $q
= 2$ would one expect, on symmetry grounds, that adding and
subtracting a vortex would give the same response.

We have also tried varying the strength of the inductive coupling
constant $\lambda_J^2$, considering $\lambda_J^2 = 0.01, 0.1$ and
$10.0$ as well as $1$ for $f=1/2$. For $\lambda_J^2 = 0.01$, the
system behaves like a collection of independent junctions: $\langle
V\rangle = 0$ for all $I/I_c < 0.9$, and the flux pattern is very
similar to that of Fig.\ \ref{flux_0p5}. $\lambda_J^2 = 0.1$ also
gives results very similar to those of Fig.\ ~\ref{flux_0p5}.   But
for $\lambda_J^2 = 10$, the fluxons are depinned even at $I/I_c =
0$, and the IV curve is very similar to that seen in Ref.\
~\cite{tornes2} for a long junction in the continuum limit.  We have
also carried out calculations (not shown here) for values of $Q_J =
1.0$, $5.0$ and $50.0$ for $f=1/2$, $N=120$ and $\lambda_J^2 = 1.0$.
The position of the first jump in the IV curve remains unchanged
even for the critically damped case ($Q_J = 1.0$). We found that,
for the larger $Q_J$'s, the fluxons are expelled from the ladder at
lower currents than for smaller $Q_J$'s.  Finally, for $Q_J = 1.0$,
the IV curve of the ladder closely resembles that of a single small,
critically damped junction.

\begin{figure}
\begin{center}
\includegraphics[width=0.45\textwidth]{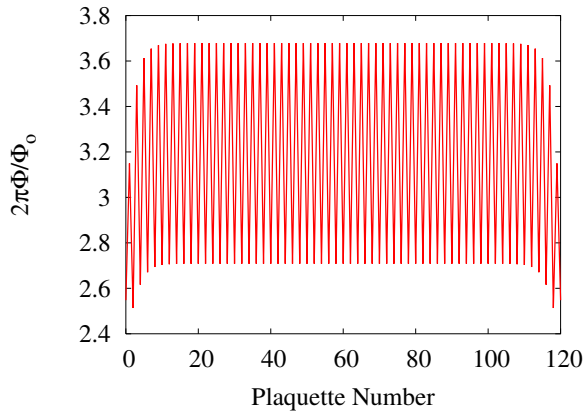}
\caption { \small Normalized flux vs.\ plaquette number, plotted for
$N=121$ and $f=61/121$.}\label{flux_odd_0p5}
\end{center}
\end{figure}

\vspace{0.05in}

\begin{figure}
\begin{center}
\includegraphics[width=0.45\textwidth]{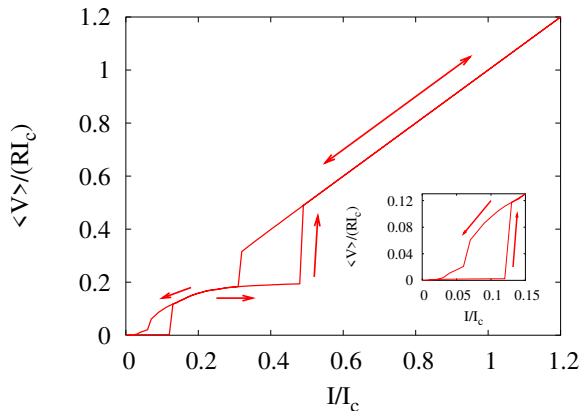}
\caption{\small IV characteristics for $N = 121$ and $f = 61/121$,
for the parameters of the preceding figure.  Inset: enlargement of
IV characteristics at small currents.}  \label{IV_odd_0p5}
\end{center}
\end{figure}

\vspace{0.05in}

\subsection{f = 1/2 + 1/(2N): One Added Half-Fluxon}

As a final calculation, we have considered a ladder with an {\em
odd} number of plaquettes, namely $N = 121$, and $f = 61/121$.
Intuitively, one might expect that this ladder would correspond to a
single added {\em 1/2-fluxon} superimposed on an $f = 1/2$ ground
state.  The latter corresponds to $f = 1/2 = 60.5/121$, while the
extra 1/2-fluxon is needed in order to bring the total flux up to $f
= 61/121$. Indeed, the flux plot, shown in Fig.\
\ref{flux_odd_0p5}, is consistent with this picture: an alternating
flux pattern of the $f = 1/2$ ground state on which is superimposed
the flux pattern of a {\em single} 1/2-fluxon, i.\ e., one of the
two excitations distinguishable in Fig.\ 6.

\begin{figure}
\begin{center}
\includegraphics[width=0.45\textwidth]{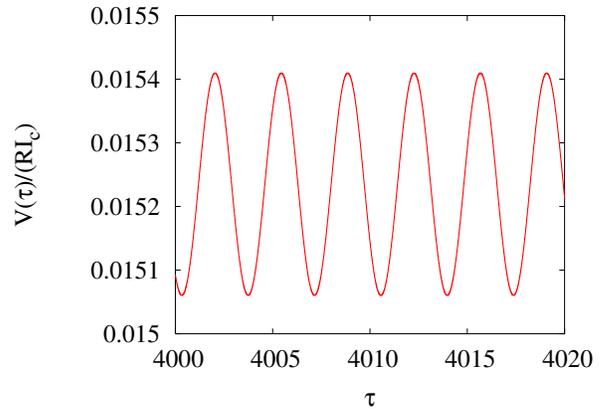}
\caption{\small Time-dependent voltage $V(\tau)/(RI_c)$ plotted for
dimensionless time $\tau$ between 4000 and 4020, for f = 61/121 and
current $I/I_c = 0.03$.  $Q_J = 10$ and $\lambda_J^2 = 1$.  The
dimensionless period of oscillation $\Delta\tau$ satisfies eq.\
(\ref{eq:period}}). \label{fig_13}
\end{center}
\end{figure}

The IV characteristics of this ladder (shown in Fig.\
\ref{IV_odd_0p5}), are consistent with this picture. There are three
regimes of non-zero voltage (seen most clearly on the decreasing
current branch), namely (i) motion of the 1/2-fluxon through the
background of the pinned $f = 1/2$ lattice; (ii) motion of the
depinned $f = 1/2$ lattice; and (iii) purely resistive behavior,
with all small junctions in the voltage state. The motion of the
half-fluxon [in regime (i)] is particularly intriguing.  At a given
current, the voltage generated by this 1/2-fluxon is $\sim 1/2$ that
produced by the two moving 1/2-fluxons at the corresponding current
in Fig.\ 6. This result is consistent with the picture that the two
1/2-fluxons in Fig.\ 6 move at the same average velocity at any
given applied current.  In addition, as mentioned earlier, at all
currents in this regime, the time-averaged voltage $\langle V
\rangle$ generated by the 1/2-fluxon is related to the ac frequency
$\nu$ of that voltage by
\begin{equation}
\frac{\langle V\rangle}{RI_c} = \frac{2\pi}{NQ_J\Delta\tau},
\label{eq:period}
\end{equation}
where $\Delta\tau = \omega_p/\nu$.
As an illustration, we show in Fig.\ \ref{fig_13} $V(\tau)/(RI_c)$
for $f = 61/121$ and a ladder of length 121 plaquettes, at a current
$I/I_c = 0.03$, in the regime where the voltage is produced by a
single 1/2-fluxon.
\vspace{0.05in}

\vspace{0.05in}

\section{Discussion}

We have shown that a single fluxon introduced into a Josephson ladder
will fractionalize if the ladder is initially in an $f = 1/q$ ground
state with $q = 2$, $3$, $4$, $5$, or $6$. This fractionalization
leads to a characteristic behavior of the (suitably defined) flux in
the ground state at $I/I_c = 0$, and even for $I/I_c \neq 0$, provided
that $I$ is smaller than the critical current for onset of
voltage. Once a measurable voltage is obtained across the ladder, for
these fields, there is a characteristic series of regimes with
increasing current: first, the fractionalized fluxon moves, generating
a voltage; then, the underlying fluxon lattice is depinned; and
finally, the ladder as a whole switches to the voltage state and the
fluxons are expelled from the ladder.

The type of fractional excitation found here are basically walls
between different domains of the underlying 1/q ground state.  For
an applied field of 1/q, there are q discrete flux lattice ground
states, each shifted by one plaquette from the next ground state. If
two different ground state domains, shifted by one plaquette, are
placed next to one another, a 1/q fractional fluxon is formed. Such
walls, with fractional vortex charge, are the analogs of fractional
excitations well known in many other physics contexts, especially in
one dimension. For example, solitons in polyacetylene are
fractionally charged excitations which can form between two domains
of alternating singly and doubly bonded carbons in this
one-dimensional chain compound\cite{ssh}.  The fractional fluxons in
the Josephson ladder are unusual because it appears that a fluxon of
{\em any} rational fraction can form in a suitable applied magnetic
field.


The fractional vortices described in this paper generally produce a
{\em non-hysteretic} IV characteristic, even if the individual
junctions which make up the ladder are hysteretic.  This behavior is
similar to that previously observed for integer vortices in
two-dimensional arrays\cite{wenbin}, which also move in a
non-hysteretic fashion even when the junctions in the array are
hysteretic.  In the case of the 2D arrays, the explanation for this
behavior is that the effective quality factor describing the voltage
generated by the moving vortex is $Q_{eff} =
(2eR^2I_dC/\hbar)^{1/2}$, where $C$ is the junction capacitance, $R$
the junction shunt resistance, and $I_d$ is the current required to
depin a vortex from its equilibrium position within a plaquette of
the 2D lattice.  Since $I_d \sim 0.1 I_c$ in the 2D case, where
$I_c$ is the critical current of one junction, $Q_{eff}$ is much
smaller than the junction quality factor $(2eR^2I_cC/\hbar)^{1/2}$,
and hence the motion of the vortex is overdamped even when the
individual junctions are underdamped and hysteretic.

For the ladder arrays, the depinning current $I_d$ for a fractional
fluxon is less than $0.01I_c$ for all cases we have considered in
this paper.  Hence, even though $Q_J = 10$, the analogous effective
quality factor for the fractional fluxon will be less than 0.1. The
resulting IV characteristics should thus be non-hysteretic and
similar to those of an overdamped Josephson junction, as we have
observed numerically (see Fig.\ 13).   However, we have not
succeeded in deriving a driven-pendulum-like equation of motion for
the fractional vortex, similar to that given in Refs.\ \cite{lobb}
and \cite{wenbin}.

One striking feature of these fractionalized fluxons is that they do
{\em not} appear to form in two dimensions.  When a fluxon is added
to an $f = 1/2$ ground state of a square lattice in two dimensions,
the fluxon is depinned at a current much lower than that of the $f =
1/2$ lattice, just as in one dimension.  However, numerical studies
suggest that the fluxon is not fractionalized, but remains a compact
object\cite{wenbin}. Fractionalization is apparently easier in 1D
because the domain walls are zero-dimensional, and hence require
much lower energy to form.

It is of interest to compare this prediction to what we would expect
from a single {\em un-fractionalized} fluxon in an otherwise empty
ladder. Here, the expected mechanism of fluxon motion is that the
fluxon moves by one plaquette.  An argument similar to that just
given yields $\langle V \rangle = [h/(2eN)\nu$.

We can also consider the motion of an un-fractionalized vortex
through the background of a pinned $f = 1/q$ fluxon lattice.  Since
the motion of the single vortex would have a period of $q$ lattice
constants, and since there would be a phase slip of $2\pi$ each time
the vortex circulates once around the ladder, the equation relating
the time-averaged voltage to the period would be $\langle
V\rangle/(RI_c) = 2\pi q/(Q_J\Delta\tau)$, where $\Delta\tau$ is the
period of the ac voltage, in units of $1/\omega_p$.  This relation
is the same as would hold for a vortex fractionalized into $q$ parts
and moving through the same background.

Thus, the clearest experimental way to see a signature of the
fractionalization would be to compare the period of the ac voltages
with the time-averaged voltage in a ladder with {\em isolated}
fractional vortices.  Such ladders were discussed earlier - an
example is the case of $N = 121$ and $f = 61/121$, which contains a
single 1/2-fluxon in a background of an $f = 1/2$ fluxon lattice.
Besides searching for a signal of the fractionalized vortices in the
IV characteristics, perhaps the flux profile of the fractionalized
vortices could also be directly imaged.


As mentioned earlier, the fractional vortices discussed here were
previously suggested\cite{chandran} for a ladder of alternating $0$
and $\pi$ junctions. Since fractional vortices can thus appear in
both types of ladders, one might ask if any information about the
nature of the junctions (i. e., whether they are $0$ or $\pi$
junctions) could be gleaned from a simple inspection of the IV
characteristics. If {\em all} the rungs of the ladder are $\pi$
junctions, they would have the same IV characteristics as a ladder
made entirely of $0$ junctions, since the presence of $\pi$
junctions would produce no additional frustration in this case.
Thus, the IV characteristics would not distinguish between all $0$
and all $\pi$ junctions.  If the ladder contained alternating $0$
and $\pi$ junctions, the IV characteristics, in the presence of a
magnetic field, would differ from that for all $0$ (or all $\pi$)
junctions by being shifted in magnetic field by 1/2 a flux quantum
per plaquette - that is, the IV characteristics of the alternating
ladder at field $f$ would be the same as that of the $0$ or $\pi$
ladder at $f + 1/2$.  For a more complicated distribution of $0$ and
$\pi$ junctions, the IV characteristics of the ladder would be more
complicated, and the presence of fractional vortices might be
difficult to detect.  Thus, the existence of fractional vortices, of
itself, might not be sufficient to determine if the ladder were
partly composed of $\pi$ junctions.


Finally, we note that the present results are all obtained from a
classical set of governing equations.  If the individual junctions
are sufficiently small, we expect that the fractional fluxons would
behave like quantum-mechanical particles.  Specifically, the phase
configuration of the ladder would be described by an N-dimensional
wave function $\Psi(\phi_1,....\phi_N)$, which would be an
eigenfunction of the ladder Hamiltonian (expressed as an operator).
It would be of interest to calculate the spectrum of such
eigenstates, and to study the behavior of a lattice, in the small
junction regime, containing two or more such quantum-mechanical
fractional fluxons.  One important question would be the statistics
(bose, fermi, or fractional) obeyed by such particles. Such
questions, though fascinating, are, however, beyond the scope of the
present paper.  In addition, there should be a broad range of
experimental parameters where the classical equations are
applicable.  Thus, the present predictions should be readily
testable experimentally.

\section{Acknowledgments}

We are grateful for valuable conversations with R.\ V.\ Kulkarni,
who called our attention to Ref.\ \cite{chandran}, and to E.\
Almaas. This work was supported by NSF Grant DMR 04-13395;
calculations were carried out using the facilities of the Ohio
Supercomputer Center.

\end{document}